\documentclass[conference]{IEEEtran}
\IEEEoverridecommandlockouts
\usepackage{subfiles}
\usepackage{cite}
\usepackage{amsmath,amssymb,amsfonts}
\usepackage{subcaption}
\usepackage{graphicx}
\usepackage{caption}
\usepackage{multirow}
\usepackage[Procedure]{algorithm}
\usepackage[noend]{algpseudocode}
\usepackage{graphicx}
\usepackage{textcomp}
\usepackage{xcolor}
\def\BibTeX{{\rm B\kern-.05em{\sc i\kern-.025em b}\kern-.08em
    T\kern-.1667em\lower.7ex\hbox{E}\kern-.125emX}}

\definecolor{orange}{rgb}{1,0.5,0}

\definecolor{cpgreen}{rgb}{0.01, 0.6, 0.24}

\usepackage{ulem}  
\usepackage{url}

\usepackage{booktabs}

\usepackage[colorinlistoftodos]{todonotes}

\begin{document}

\title{Benchmarking the Linear Algebra Awareness of TensorFlow and PyTorch\\
\thanks{Financial support from the Deutsche Forschungsgemeinschaft (German Research Foundation) through grants GSC 111 and IRTG 2379 is gratefully acknowledged.}
}



\author{
	
	\IEEEauthorblockN{Aravind Sankaran\IEEEauthorrefmark{1},
Navid Akbari Alashti\IEEEauthorrefmark{2}, and Christos Psarras\IEEEauthorrefmark{3}}
	
	\IEEEauthorblockA{RWTH Aachen University, Germany \\
	\{\IEEEauthorrefmark{1}aravind.sankaran, \IEEEauthorrefmark{2}navid.alashti, \IEEEauthorrefmark{3}christos.psarras\}@rwth-aachen.de \\
}
			\and
	\IEEEauthorblockN{Paolo Bientinesi}
	\IEEEauthorblockA{Ume\r{a} Universitet, Sweden \\
	pauldj@cs.umu.se}
}

\maketitle
\thispagestyle{plain}
\pagestyle{plain}


\begin{abstract}
	Linear algebra operations, which are ubiquitous in machine learning, form major performance bottlenecks. 
	The High-Performance Computing community invests significant effort in the development of architecture-specific optimized kernels, such as those provided by the BLAS and LAPACK libraries, to speed up linear algebra operations. 
	However, end users are progressively less likely to go through the error prone and time-consuming process of directly using said kernels; instead, frameworks such as TensorFlow (TF) and PyTorch (PyT), which facilitate the development of machine learning applications, are becoming more and more popular.  
	Although such frameworks link to BLAS and LAPACK, it is not clear whether or not they make use of linear algebra knowledge to speed up computations. 
	For this reason, in this paper we develop benchmarks to investigate the linear algebra optimization capabilities of TF and PyT. 
	Our analyses reveal that a number of linear algebra optimizations are still missing; for instance, reducing the number of scalar operations by applying the distributive law, and automatically identifying the optimal parenthesization of a matrix chain. 
	In this work, we focus on linear algebra computations in TF and PyT; we both expose opportunities for performance enhancement to the benefit of the developers of the frameworks and provide end users with guidelines on how to achieve performance gains.

\end{abstract}

\section{Introduction}
\label{sec:introduction}

The recent advances to machine learning technologies have led to the development of many latency sensitive applications, such as intelligent vehicles~\cite{grewe2017information}, real-time video analytics~\cite{ananthanarayanan2017real}, and augmented reality~\cite{schmoll2018demonstration}, that require fast execution of costly scientific computations on heterogeneous hardware.
One of the major performance bottlenecks for such scientific computations is the evaluation of linear algebra expressions, which are building blocks for countless computational problems. 
Libraries such as BLAS and LAPACK  provide a small set of high performance kernels for some standard operations that are used to build linear algebra expressions.
Implementing a linear algebra expression by directly calling said kernels is a laborious task, time consuming even for
domain experts, as they should familiarize themselves with the syntax of those calls, along with its input parameters,
data-types and storage formats.

By contrast, frameworks such as TensorFlow~\cite{abadi2016tensorflow}, PyTorch~\cite{paszke2019pytorch},
Matlab~\cite{matlab}, Julia~\cite{bezanson2017julia}, Eigen~\cite{guennebaud2010eigen},
Armadillo~\cite{sanderson2016armadillo} (as well as many others) allow end users to input a linear algebra expression at a high level of abstraction, where the syntax closely resembles the one used on a blackboard. 
Internally, these tools compile the expression by automatically mapping the users' input to a sequence of function calls to basic operations that are computed with the optimized libraries. 
However, the mapping of a high-level expression to an optimized sequence of library calls is a task far from trivial; because, a linear algebra expression can be computed in many different ways---each corresponding to a specific sequence of
library calls---which can significantly differ in terms of performance from one another~\cite{barthels2021linnea, sankaran2021discriminating}.
Unfortunately, it has been found that the mapping done by most popular high-level programming languages is still sub-optimal~\cite{psarras2021linear}. 
In this paper, we extend this investigation to popular machine learning frameworks: TensorFlow and PyTorch.

\begin{figure}
	\includegraphics[width=0.45\textwidth]{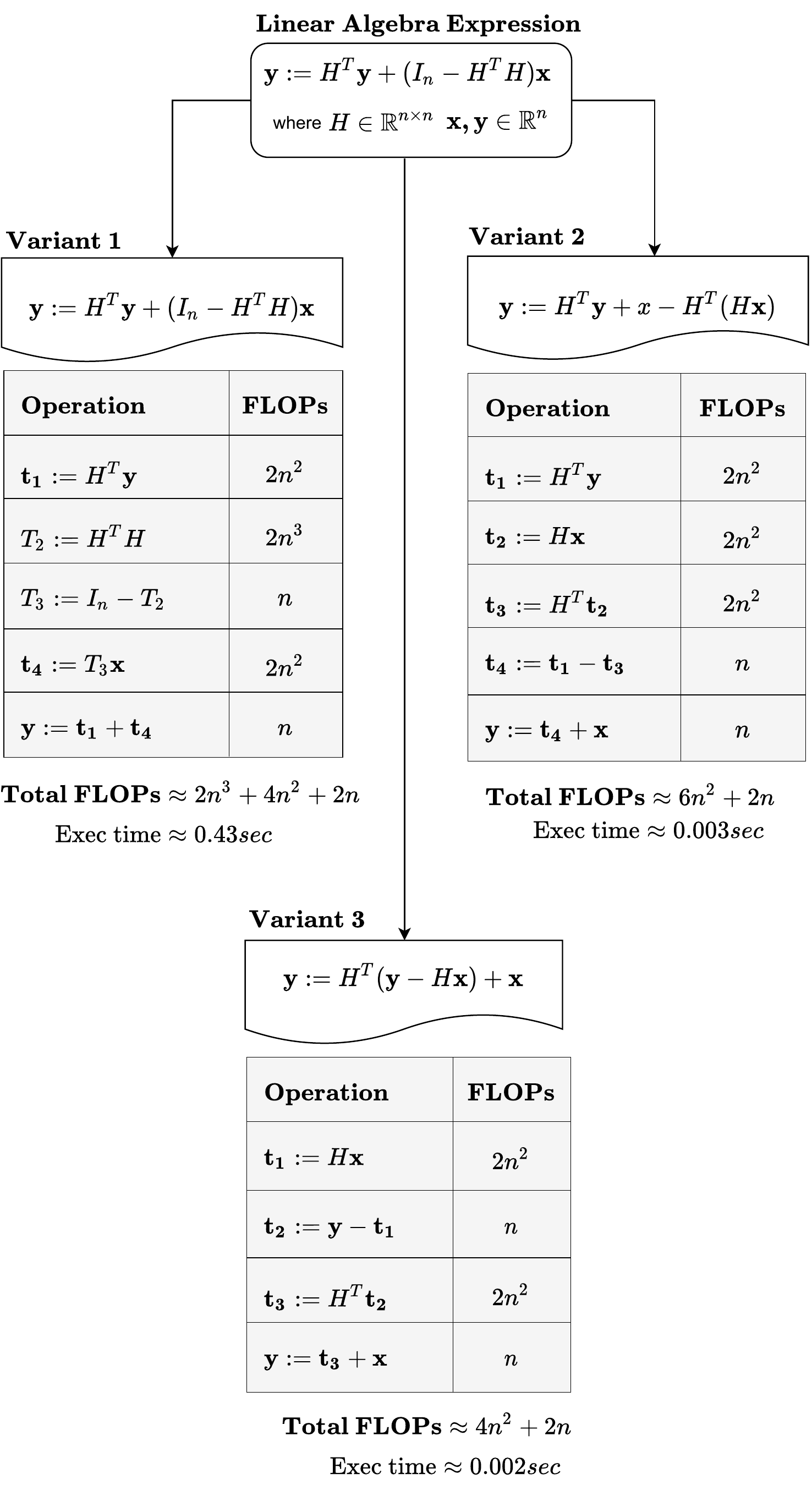}
	\caption{Variants of Image restoration equation. The execution time is reported for $n=3000$.}
	\label{fig:img-rest}
\end{figure}

The knowledge of linear algebra can be used to derive variants for computing a given input expression;
for instance, consider the following expression which appears in an image restoration application~\cite{tirer2018image}: 
\begin{equation}
\label{eqn:img-rest}
\mathbf{y} := H^{T} \mathbf{y} + (I_{n} - H^{T} H )\mathbf{x}
\end{equation}
where $H$ is a square matrix in $\mathbb{R}^{n \times n}$ , $\mathbf{x}$ and $\mathbf{y}$ are vectors in $\mathbb{R}^{n}$, and $I_{n}$ is the identity matrix. 
This expression can be rewritten in several ways by applying the distributive and associative properties; the variants are mathematically equivalent but the sequences of instructions are different (see in Fig \ref{fig:img-rest}).
The domain scientists may prefer to work with Variant~1 as it is more descriptive of the underlying physics. 
However, in terms of performance, Variant~1 would be a suboptimal choice because the product $H^{T}H$ is computed explicitly, performing an expensive  $\mathcal{O}(n^3)$ matrix-matrix multiplication. 
By applying distributivity, the term $ (I_{n} - H^{T} H )\mathbf{x}$ in Expression~\ref{eqn:img-rest} is expanded, yielding:
\begin{equation}
\label{eqn:img-rest-var2}
\mathbf{y} := H^{T} \mathbf{y} + \mathbf{x} - H^{T}H\mathbf{x}
\end{equation}
Due to the associativity of matrix products, the term $ H^{T}H\mathbf{x}$ in Expression~\ref{eqn:img-rest-var2} can either be evaluated from ``left-to-right'': $ (H^{T}H)\mathbf{x}$  or ``right-to-left'': $H^{T}(H\mathbf{x})$. 
The left-to-right parenthesization  does not get rid of the expensive $\mathcal{O}(n^3)$ matrix multiplication. 
However, the right-to-left parenthesization lowers the complexity of Expression~\ref{eqn:img-rest-var2}, which now consist of three matrix-vector products, to $\mathcal{O}(n^2)$. 
Furthermore, applying the distributive property again, Expression~\ref{eqn:img-rest-var2} can be re-written as:
\begin{equation}
\label{eqn:img-rest-var3}
\mathbf{y} := H^{T} (\mathbf{y} - H\mathbf{x}) + \mathbf{x}
\end{equation}
Now, the total number of FLOPs is further reduced, as there are only two matrix-vector products. 
The average execution time of the three variants implemented in TensorFlow, linked to Intel-MKL on an Intel Xeon CPU\footnote{for further details on the hardware used, see Sec~\ref{sec:eval}}, is shown in Fig~\ref{fig:img-rest}. 
The execution of variant 2 and 3, which does not consist of a matrix matrix product is significantly faster than variant 1.

\textbf{\textit{Contributions}}: In this paper, we consider the two most popular machine learning frameworks TensorFlow and PyTorch, and investigate the extent to which these frameworks incorporate linear algebra knowledge to optimize mathematical expressions. 
To this end, we develop benchmarks consisting of simple tests that expose whether or not the frameworks carry out linear algebra optimizations such as exploiting the knowledge of matrix properties (e.g., triangular, symmetric, etc.) to reduce the number of FLOPs, optimal parenthesization of matrix chains, code motion, etc. 
Our benchmarks not only serve as a guide for end users to help them achieve better performance, but also expose optimization opportunities to the developers of these frameworks to help improve performance when computing linear algebra expressions.

\paragraph*{\textbf{Organization}} In Sec.~\ref{sec:rel}, we survey the state of art. In Sec.~\ref{sec:eval}, we describe our experiments, evaluate the frameworks and discuss our findings. Finally, in Sec~\ref{sec:conclusion}, we draw conclusions.

\section{Related Works}
\label{sec:rel}

The difficulty in deploying and tuning the performance of scientific programs on diverse hardware and operating systems has boosted the research and development of compilers that have multiples levels of Intermediate Representations (IRs). 
Modern compilers such as TVM~\cite{chen2018tvm}, Intel nGraph~\cite{cyphers2018intel},  XLA~\cite{leary2017xla}, Tensor Comprehension~\cite{vasilache2018tensor} and Glow~\cite{rotem2018glow} translate a mathematical problem through multiple levels of Intermediate Representations (IRs)---while optimising them for performance---and eventually generate machine code.
Depending on the hardware supported by the compiler, machine learning  frameworks such as TensorFlow~\cite{abadi2016tensorflow}, PyTorch~\cite{paszke2019pytorch}, MXNet~\cite{chen2015mxnet} and CNTK~\cite{seide2016cntk} are built with one of these compilers at their backend.
Multi-level IR modularizes compiler development, so that at each level, a specific class of optimizations can be carried out~\cite{li2020deep}.
At the top level, the input problem is cast to a computational graph, which captures the dependencies between data and the mathematical operations; here, algebraic optimizations such as common sub-expression elimination to remove redundant computations, algebraic simplifications, loop invariant code motion, etc., are carried out.
The optimized graph is then passed on to third-part tools such as LLVM~\cite{lattner2004llvm} for translation to lower-level IRs, where hardware-specific optimizations (such as instruction scheduling, static memory allocation, etc.)  are carried out, and the code is generated for the target architecture. 

The compilation of an input linear algebra expression to generate an optimized implementation can be seen as a translation to a sequence of optimized library calls. 
Libraries such as Intel's MKL and oneDNN~\cite{oneDNN}, NVIDIA's cuBLAS and cuDNN~\cite{chetlur2014cudnn},  BLIS~\cite{van2015blis} etc. encapsulate operations that occur frequently in machine learning and linear algebra, and provide hardware-optimized kernels. 
Each expression can be computed by potentially hundreds of alternative sequences that are mathematically equivalent but significantly differ from one another in terms of performance~\cite{barthels2021linnea,sankaranPAISE}.
The problem of mapping a target expression into a sequence of optimized calls, while minimizing a performance-based cost function is known as the Linear Algebra Mapping Problem (LAMP); 
the solution to LAMP is at least NP-Complete and it has been found that many popular frameworks for high-level languages~\cite{matlab,bezanson2017julia,guennebaud2010eigen,sanderson2016armadillo} provide sub-optimal solutions~\cite{psarras2021linear}. 

LAMP can be tackled at the compilers' top level IR via graph optimization techniques. 
DxT~\cite{marker2013code} encodes linear algebra programs as directed acyclic graphs and uses a greedy approach to transforms them into sequence of high-performing library routines targeting distributed memory architectures. 
Linnea~\cite{barthels2021linnea} is a linear algebra compiler that uses graph based approach to automatically generate variants for a given input expression and recommends an optimal variant based on the number of FLOPs performed.
Grappler~\cite{larsen2019tensorflow} is TensorFlows graph optimization engine that targets common machine learning operations. 
Since both TensorFlow and PyTorch are known to be able to perform a number of graph optimizations at their backends~\cite{li2020deep}, in this paper we develop benchmarks that expose further opportunities for improving their performance when it comes to linear algebra computations.

While there are already several machine learning benchmarks such as~\cite{reddi2020mlperf,BaiduDeepBench,coleman2017dawnbench}, etc., which enable users to compare different frameworks and choose a more suitable one to implement their ideas with, none of them---to the best of our knowledge---evaluate the performance of linear algebra expressions within the frameworks~\cite{DLBenchmarksSurvey}.
In this paper, we follow a similar approach to~\cite{psarras2021linear} to investigate the performance of linear algebra expressions executed within TensorFlow and PyTorch.

\section{Evaluation of the Frameworks}
\label{sec:eval}

In our experiments, we input linear algebra test expressions in a mathematically descriptive syntax using the Python interface, and evaluate the solutions provided by TensorFlow (TF) and PyTorch (PyT)  in terms of performance. 
The frameworks translate the python code into a language-independent intermediate representation, where the input expression is broken down into basic mathematical operations that are executed using optimized kernels.
For a given test expression, we assess the quality of the solution by comparing it with several other optimized, but mathematically equivalent alternatives; the alternatives typically differ in terms of the required number of FLOPs or memory accesses. 
In order to compare the performance, we measure the single-threaded execution time and report the minimum over 20 repetitions; we check whether the performance differences are statistically significant (or not) using the boot-strapping approach from~\cite{sankaran2021discriminating}. 
The experiments are run using a single core on a Linux machine with an Intel Xeon Platinum 8160 processor.
The versions of the languages used are the latest stable releases as of December 2021: TensorFlow 2.7.0 and PyTorch 1.10.1.
The frameworks are interfaced via Python 3.9.7 and linked to the Math Kernel Library (MKL) through the Intel OneAPI interface 2022.0.0, which provides the optimized kernels for Intel architectures. 
The source code for the experiments is available online\footnote{https://github.com/as641651/LinearAlgebra-Awareness-Benchmark}.

The development of machine learning models typically involves two phases: Training and Deployment. 
The training is carried out in a research and development setting, where fast prototyping and experimentation often become more important than performance. 
During the deployment phase instead, the models are executed in a production environment that might demand low latency and high performance.  
In order to improve the productivity of developing machine learning applications and minimize code modifications between the training and deployment phases, TF and PyT support two programming modes: ``Eager''  and ``Graph''.  The Eager mode is intended for research and development, while Graph mode is intended for production use cases.

\subsection*{\textbf{Graph mode vs Eager mode}}
In the Graph mode, the structure of the user code is used to capture the control flow of data and operations, and to generate a computational graph, over which rigorous performance optimizations are carried out. 
In the Eager mode, performance optimizations are not as extensive as those in the Graph mode.
\begin{figure*}
	\begin{minipage}{.45\linewidth}
		\includegraphics[width=0.9\textwidth]{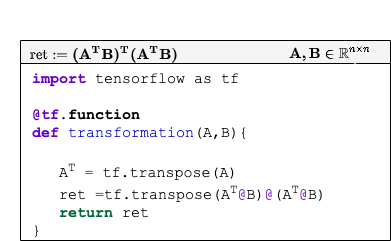}
		\caption{TensorFlow Code.}
		\label{fig:newton-code}
		
		\vspace*{\floatsep}
		\begin{center}
			\renewcommand{\arraystretch}{1.2}
			\begin{tabular}{@{}l ccc@{}}
				\toprule
				Expression & MKL-C   & Eager & Graph \\
				&&\small{(TF / PyT)}&\small{(TF / PyT)}\\
				\midrule
				{$A^TB$ \hfill }& 0.39  & 0.40 /  0.40 & 0.40 / 0.40 \\
				{$(A^TB)^T(A^TB)$ } & --  & 1.25 / 1.27   & 0.78 / 0.80 \\
				\bottomrule
			\end{tabular}
		\end{center}
		\captionof{table}{Execution time (in sec) for $n = 3000$}
		\label{tab:cse}
	\end{minipage}
	\begin{minipage}{.55\linewidth}
		\includegraphics[width=\textwidth]{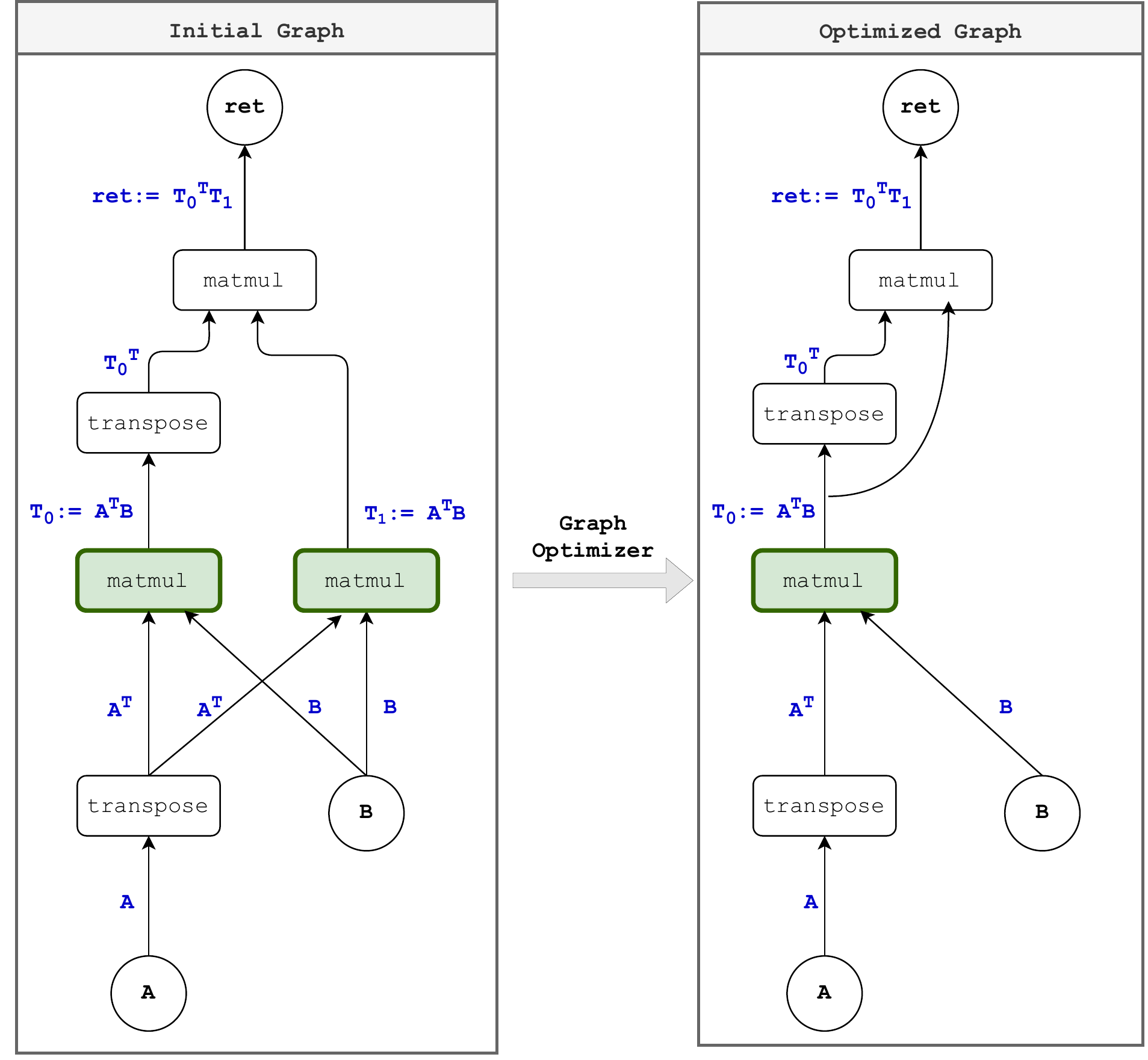}
		\caption{The Computational Graphs for $(A^TB)^T(A^TB)$.}
		\label{fig:newton-graph}
	\end{minipage}
	
\end{figure*}

For illustration, consider the following expression which occurs in the Stochastic Newton equations~\cite{chung2017stochastic}:
\begin{equation}
\label{eqn:sn2}
Y := (A^TB)^T(A^{T}B) \qquad A,B \in \mathbb{R}^{n \times n}
\end{equation}
The TF code in Python that computes Expression~\ref{eqn:sn2} is shown in Figure~\ref{fig:newton-code}.  
The operator ``$@$'' performs the matrix-matrix multiplication operation.
By default, the code executes in Eager mode. 
In order to execute the code in Graph mode, the TF function has to
be wrapped using the decorator \textbf{@tensorflow.function} (PyT functions have to be wrapped with \textbf{@torch.jit.scripts}).
Unlike Eager mode, the loops inside the wrapped functions of the Graph mode cannot be programmed with the usual pythonic syntax, but have to be handled specially using framework specific constructs. 
This is due to the fact that, in the Graph mode, the user code is translated into a Directed-Acyclic-Graph (DAG), which does not contain loops or cycles; the nodes of the graph represent basic operations such as matrix product, transpose, etc., and the edges represent data flow.

In both Eager and Graph mode, the frameworks link to MKL, which offers highly optimized kernels for frequently encountered linear algebra operations. 
In order to confirm, we compare the performance of matrix matrix multiplication with a reference implementation that use the MKL kernel for matrix multiplication, GEMM, implemented in C.
In the first row of Table~\ref{tab:cse}, we report the execution time for the matrix product sub-expression from Expression~\ref{eqn:sn2}  for $n=3000$ on single precision floating point data\footnote{As many ML applications operate on
	single precision data, both TF and PyT use the single precision floating point representation by
	default.} in C, TF and PyT. We observe no statistically significant difference in timings (after ignoring overheads for loading the decorators\footnote{@tf.function = 6e-4 sec and @torch.jit,script = 2e-3 sec}) between the reference implementation and the frameworks in both Graph and Eager mode; hence, we confirm that the frameworks do link to MKL.

In the Graph mode, a number of graph-optimizations are applied over the DAG; for example, duplicate nodes that compute the exact same operation for the same input data can be identified and removed.
The non-optimized and optimized DAG for the code-snippet in Fig.~\ref{fig:newton-code} is shown in
Fig.~\ref{fig:newton-graph}; the circular nodes represent I/O operations, while the rounded-rectangle nodes represent
mathematical operations.
In the non-optimized graph, the matrix multiplication or the \textbf{matmul} operation $A^TB$ is computed twice;
removing one of the nodes saves $2n^3$ FLOPs. Such optimizations are not carried out when the code is executed in Eager
mode (i.e., without the decorators). 
In the second row of Table~\ref{tab:cse}, we report the execution times in the Eager and the Graph mode; the execution time in the Eager mode is approximately 1.5 times longer than in the Graph mode.

In the following, we  consider only the Graph mode and develop experiments to investigate further optimization opportunities in the computation of linear algebra expressions and benchmark the performance impacts.  To this end, we developed a set of 5 experiments, each one of them containing one or more test expressions, to identify whether or not the frameworks perform a particular linear algebra optimization.

\begin{figure*}
	\begin{minipage}{.35\linewidth}
		\begin{center}
			\renewcommand{\arraystretch}{1.2}
			\begin{tabular}{@{}ll ll@{}}
				\toprule
				&Expression& TF & PyT \\
				\midrule
				1&{$A^TB$ \hfill }& 0.40  & 0.40  \\
				2&{$A^TB + A^TB$ } & 0.40  & 0.41   \\
				\cmidrule{1-4} 
				3&{$(A^TB)^T(A^TB)$ \hfill }& 0.78  & 0.80  \\
				4&{$(A^TB)^TA^TB$ } & 1.17  & 1.15   \\
				\bottomrule
			\end{tabular}
		\end{center}
		\captionof{table}{Execution time (in sec) for \\ $n = 3000$}
		\label{tab:exp1}
	\end{minipage}
	\begin{minipage}{.65\linewidth}
		\includegraphics[width=\textwidth]{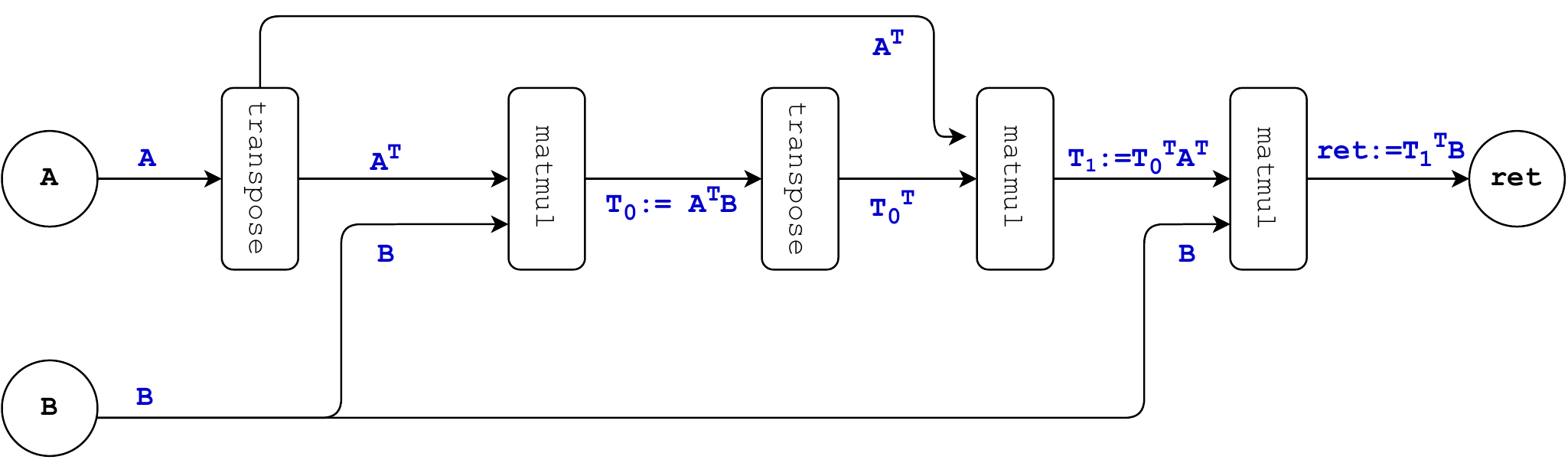}
		\caption{The Computational Graph for $(A^TB)^TA^TB$.}
		\label{fig:newton-matmul}
	\end{minipage}
\end{figure*}

\subsection{\textbf{Experiment 1: Common Sub-expression Elimination }}
Common Sub-expression Elimination (CSE) is a typical optimization incorporated in most modern compilers to reduce
redundant computations. 
To this end, sub-expressions within an expression that evaluate to the same value are computed only once, and substituted through a temporary reference in all the subsequent instances.
As illustrated earlier in this section, both TF and PyT detect common sub-expressions by searching for duplicate operations in a DAG.
In this experiment, we investigate the extent to which the DAG-based approach can successfully detect common sub-expressions in some of the typical high-level input expressions.

\textbf{\textit{Test expressions: }} We consider the following sub-expression $S$ that occurs in
the Stochastic Newton equations~\cite{chung2017stochastic}: 
\begin{equation*}
\label{eq:cse1}
S := A^TB \qquad A,B \in \mathbb{R}^{n \times n}
\end{equation*}
We develop two kinds of test expressions in which $S$ occurs multiple times, via summation and multiplication. 
The test expressions and their execution times for $n=3000$ are shown in Table~\ref{tab:exp1}\footnote{In all the following experiments, the problem size is $n=3000$, unless specified otherwise.}.
\begin{itemize}
	\setlength\itemsep{1em}
	\item \textbf{\textit{Repeated in summation:}} In this expression, $S$ occurs 
	twice, and the occurrences are added together:
	\begin{equation*}
	\label{eq:sum-cse}
	E_1 := A^TB + A^TB.
	\end{equation*}
	We compare the execution time to compute $E_1$ with that for $S$. 
	Without CSE, the execution time for $E_1$ would be approximately $2\times$ higher than that for $S$.
	\paragraph*{\textbf{Result}} In both frameworks, the execution time for $E_1$ is close to that for $S$.
	Therefore, we infer that the frameworks successfully identified that the $S$ appears twice in $E_1$, and avoided redundant computation by instead performing $2A^TB$.
	The complexity of scaling $S$ by 2 is only $\mathcal{O}(n^2)$, whose execution time is negligible in comparison to the $\mathcal{O}(n^3)$ matrix product. 
	Moreover, in the BLAS specification, the scaling does not have to occur separately, but can be done alongside multiplication without additional overheads.
	
	\item \textbf{\textit{Repeated in multiplication:}} Let us now consider the expression $B^TAA^TB$, which a user can input in many different ways. 
	Let us assume that the user explicitly rewrites the expression by distributing the transposition\footnote{$U^TV^T = (VU)^T$} so that $S$ occurs twice; the input can then be either $E_2$ or $E_3$:
	\begin{equation*}
	\label{eq:mul-cse}
	\begin{aligned}
	E_2 :=  &  (A^TB)^T(A^TB) \qquad &\text{(parenthesized)} \\ 
	E_3 :=  &  (A^TB)^TA^TB \qquad &\text{(non-parenthesized)} 
	\end{aligned}
	\end{equation*}
	With CSE, the input expressions would consist of only two matrix multiplications -- one to compute $S$, and the other to compute $S^TS$ -- and the execution time would be approximately $2\times$ higher than that for $S$. 
	Without optimization, there would be additional matrix multiplication, and the execution time would be approximately $3\times$ higher than that for $S$.
	\paragraph*{\textbf{Result}} In both frameworks, the execution time of the parenthesized expression $E_2$ is $2\times$ the time for $S$, indicating that the common sub-expression was detected and the redundant operation was avoided.
	However, the execution time of the non-parenthesized expression $E_3$ is close to $3\times$ the time for $S$, 
	which shows that if the user does not explicitly indicate the parenthesization, the frameworks fail to detect the common sub-expression.
\end{itemize}

\textbf{\textit{Discussion:}} In-order to detect a common sub-expression, the DAG should consist of duplicate nodes with
similar inputs.
The graph generated by TF for the non-parenthesized matrix chain expression $(A^TB)^TA^TB$ is shown in Figure~\ref{fig:newton-matmul}.
The graph does not consist of any duplicate nodes.
This observation shows that the frameworks do not attempt to find the optimal parenthesization for our given matrix chain input, which leads us to the next experiment, where we investigate the consequences of non-optimal parethesization of matrix chains.

\subsection{\textbf{Experiment 2: Optimization of Matrix Chains}}

Given $m$ matrices of suitable size, the product $M := A_1 A_2 . . . A_m$ is known as a matrix chain. Because of
associativity of the matrix product, matrix chains can be computed in many different ways,
each identified by a specific parenthesization; for instance, $M := A_1A_2A_3 $ can be computed either as $(A_1A_2)A_3$
or $A_1(A_2A_3)$. The number of different parenthesizations for a chain of length $m$ is given by the $(m-1)^{\text{th}}$ Catalan number:
\begin{equation*}
C_{m-1} = \frac{(2m)!}{(m+1)!m!}.
\end{equation*}
Although different parenthesizations evaluate to the same mathematical result (they all compute $M$), they require different number of FLOPs,
and hence exhibit different performance.
In this experiment, we input matrix chains without any parenthesization, and investigate if the frameworks automatically choose an evaluation order that requires minimum FLOPs.
\begin{table}[h!]
	\normalsize
	\begin{center}
		\renewcommand{\arraystretch}{1.2}
		\begin{tabular}{@{}l l c ll@{}}
			\toprule
			Expression & \multicolumn{1}{c}{TF} && \multicolumn{2}{c}{PyT} \\
			\cmidrule{2-2} \cmidrule{4-5}
			& matmul && matmul & multi\_dot \\
			\cmidrule{1-5}
			{$H^TH\mathbf{x}$ \hfill }& 0.40  && 0.41 & 0.006 \\
			{$H^T(H\mathbf{x})$} & 0.006  && 0.004 & -  \\
			\cmidrule{1-5} 
			{$\mathbf{y}^TH^TH$ \hfill }& 0.006  && 0.005 & 0.005 \\
			{$(\mathbf{y}^TH^T)H$} & 0.006  && 0.005 & -  \\
			\cmidrule{1-5} 
			{$H^{T}\mathbf{y}\mathbf{x}^{T}H $} & 0.41   && 0.40 & 0.01  \\
			{$(H^{T}\mathbf{y})(\mathbf{x}^{T}H) $} & 0.01   && 0.01 & - \\
			\bottomrule
		\end{tabular}
		\caption{Execution time (in sec)  of matrix chain test expression for $n=3000$. }
		\label{tab:exp2}
	\end{center}
\end{table}

\textbf{\textit{Test expressions:}}  We consider three different matrix chains with sizes of the matrices chosen so that
the evaluation order that minimizes the number of FLOPs is attained by evaluating the chain from \textit{right-to-left} $A_1(A_2A_3)$, \textit{left-to-right} $(A_1A_2)A_3$, and
in \textit{mixed} order $(A_1A_2)(A_3A_4)$:
\begin{itemize}
	\setlength\itemsep{1em}
	\item \textbf{\textit{Right-to-Left:}} Recall the matrix chain sub-expression from the image restoration problem (Expression~\ref{eqn:img-rest}):
	\begin{equation}
	\label{eq:matchain}
	\hat{\mathbf{y}_1} : = H^TH\mathbf{x} \qquad H \in \mathbb{R}^{n \times n}, \quad \mathbf{x} \in \mathbb{R}^{n }
	\end{equation}
	The right-most operand in the above expression is a vector; as a consequence, the evaluation from \textit{right-to-left} $H^T(H\mathbf{x})$ avoids the expensive $\mathcal{O}(n^3)$ matrix product, and reduces the complexity to $\mathcal{O}(n^2)$, incurring a cost of $2n^2 + 2n^2$ FLOPs.  On the other hand, since $H$ is a matrix, the evaluation from left-to-right $(H^TH)\mathbf{x}$ is $\mathcal{O}(n^3)$, costing $2n^3 + 2n^2$ FLOPs. We input the expression without explicit parenthesization to check whether or not the frameworks automatically choose the \textit{right-to-left} evaluation order.
	\paragraph*{\textbf{Result}} The execution times to compute the test expressions are shown in
	Table~\ref{tab:exp2}. In both frameworks, when the syntax of the matrix chain is expressed using the regular
	matrix multiplication operator ``@'' (matmul), the execution time for the expression without parenthesization is
	significantly greater than that for the same expression evaluated from \textit{right-to-left}, which shows that
	the frameworks do not automatically choose the optimum parenthesization.
	
	\item \textbf{\textit{Left-to-Right:}} Consider the following expression:
	\begin{equation}
	\hat{\mathbf{y}_2} := \mathbf{y}^TH^TH \qquad H \in \mathbb{R}^{n \times n}, \quad \mathbf{y} \in \mathbb{R}^{n \times 1}
	\end{equation}
	Since the vector $\mathbf{y}$  is appears in the left end of the matrix chain, the
	evaluation from \textit{left-to-right} $(\mathbf{y}^TH^T)H$ is $\mathcal{O}(n^2)$, whose complexity is one order
	of magnitude lower than the evaluation from \textit{right-to-left} $\mathbf{y}^T(H^TH)$  that consists of an $\mathcal{O}(n^3)$ matrix product.
	\paragraph*{\textbf{Result}} In both frameworks, the execution time for the test expression expressed using the matmul operator without parenthesization is comparable to that of the same expression evaluated from \textit{left-to-right}. Therefore, we infer that the default evaluation of matrix chains in both the frameworks is from \textit{left-to-right}.
	
	\item \textbf{\textit{Mixed:}} Consider the following expression,
	\begin{equation}
	\hat{Y} := H^{T}\mathbf{y}\mathbf{x}^{T}H \qquad H \in \mathbb{R}^{n \times n}, \quad \mathbf{x,y} \in \mathbb{R}^{n \times 1}
	\end{equation}
	where the vectors $\mathbf{x,y}$ occur at the middle of a matrix chain, and neither \textit{left-to-right} nor \textit{right-to-left} evaluation avoids the $\mathcal{O}(n^3)$ matrix matrix product; instead, the evaluation $(H^{T}\mathbf{y})(\mathbf{x}^{T}H)$ turns out to be the optimum, having $\mathcal{O}(n^2)$ complexity.
	\paragraph*{\textbf{Result}} The execution time for the test expression without parenthesisation is significantly
	more than that for the expression evaluated in the optimized order; as usual, the frameworks do not automatically
	identify the optimum evaluation order when the input is expressed without explicit parenthesization using the matmul operator.
\end{itemize}
\begin{figure}[h!]
	\begin{center}
		\includegraphics[width=0.45\textwidth]{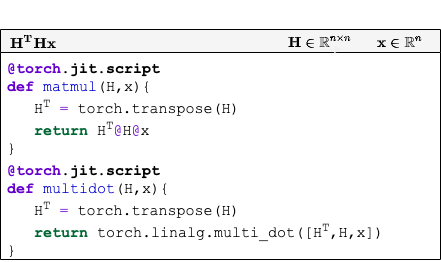}
		\caption{Variants of Matrix Chain.}
		\label{fig:multidot}
	\end{center}
\end{figure}

\begin{figure}
	\includegraphics[width=0.45\textwidth]{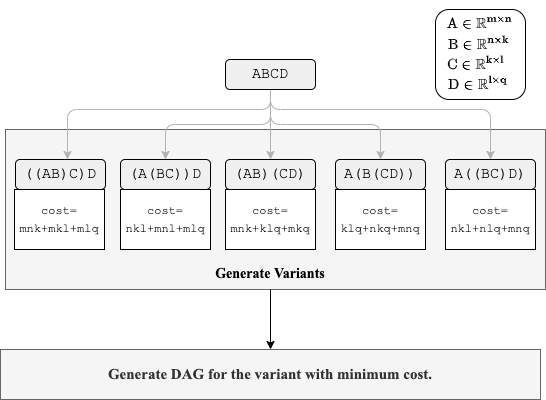}
	\caption{Variants for Matrix Chain of length 4.}
	\label{fig:img-matchain}
\end{figure}

\textbf{\textit{Discussion:}}  Notice that PyTorch offers also the \textbf{torch.linalg.multi\_dot} method, which solves the matrix chain problem and automatically selects the parenthesization that requires minimum FLOPs. The syntax of the input expression using the \textbf{multi\_dot} function is shown in the code-snippet in Figure~\ref{fig:multidot}. The optimum evaluation order in terms of \#FLOPS can be found using dynamic programming, where all possible parenthesizations are listed, and the variant with minimum FLOPs is chosen for execution; for a matrix chain of length 4, we show the different parenthesizations and their associated FLOPs in Figure~\ref{fig:img-matchain}. However, it should be noted that variants with comparable FLOP counts do not always exhibit comparable execution time. For instance, the evaluation $(AB)(CD)$ can correspond to two different implementations as shown in the code-snippet in Figure~\ref{fig:cache}, which differ in the order of instructions: 
\begin{center}
	\includegraphics[width=0.45\textwidth]{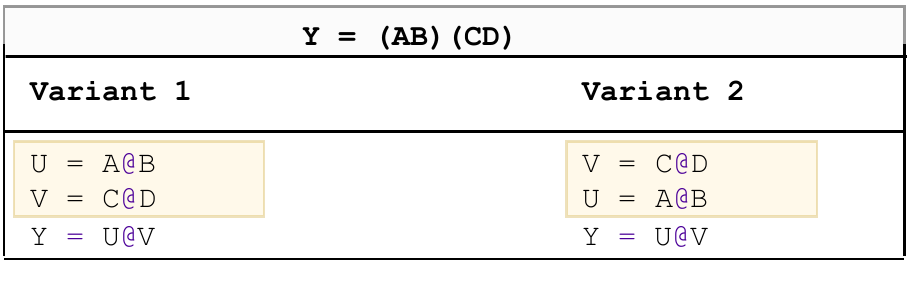}
	\captionof{figure}{Variants that require same FLOPs. }
	\label{fig:cache}
\end{center}
Although both the variants require the same FLOP count, their execution times might differ due to the differences in memory overheads~\cite{peise2014study}. Moreover, minimizing FLOP count does not always minimize execution time, especially when the overheads due to memory references dominate; increasing parallelism tend to shift the problem to be memory-bound\footnote{The performance is limited by the speed of memory access.}. However, many linear algebra computations with dense matrices (having only a few non-zero elements) execute in the compute bound\footnote{The performance is limited by the speed of the CPU.}, hence FLOPs could still be a significant indicator of performance for the experiments considered in this paper~\cite{barthels2021linnea}.

The number of FLOPs required to perform a regular matrix-matrix multiplication (GEMM) is well known, hence the performance cost can be approximated using the formulae in Figure~\ref{fig:img-matchain}. However, when the matrices in multiplication have special properties such as being Symmetric, Triangular, etc, BLAS offers specialized kernels for matrix product that require fewer FLOPs than GEMM; if such specialized kernels are used, the cost function should be modified according to the FLOPs computed by that kernel. In the following experiment, we investigate if the frameworks take advantage of those specialized kernels to speed up linear algebra operations. 

\subsection{\textbf{Experiment 3: Exploiting Matrix Properties}}

Matrix multiplications with matrices having certain special properties can be accelerated using specialized BLAS kernels; for instance, if one of the matrices in the multiplication is triangular, the product can be computed using the kernel TRMM for half the number of FLOPs than GEMM. 
In Python, these kernels can be accessed directly through the interface provided by the SciPy library~\cite{virtanen2020scipy}, which allows the end users to code the mathematical problem by explicitly calling said kernels; however, this task is time consuming and laborious as the users should familiarize themselves with the syntax of those calls. 
In order to ease code development, both TF and PyT encapsulate the direct access to these kernels and offer a mathematically descriptive interface, however, it is then not clear whether or not the frameworks make use of the knowledge of matrix properties to speed up computations.
Therefore, in this experiment, we compare the execution time of the hand-coded SciPy implementations -- specifically written to make use of specialized kernels --, with the execution time of TF/PyT for those tests, to determine whether or not the frameworks offer  similar provisions that exploit matrix properties to speed up computations.

\begin{table}[h!]
	\normalsize
	\begin{center}
		\renewcommand{\arraystretch}{1.2}
		\begin{tabular}{@{}c c c cc c cc@{}}
			\toprule
			Expr & \multicolumn{1}{c}{SciPy} &&\multicolumn{2}{c}{TF} && \multicolumn{2}{c}{PyT} \\
			\cmidrule{2-2} \cmidrule{4-5} \cmidrule{7-8}
			& BLAS && matmul & optim && matmul & optim \\
			\cmidrule{1-8}
			{$AB$\hfill }& 0.40 && 0.40  & - && 0.41 & - \\
			\cmidrule{1-8} 
			{$LB$\hfill }& 0.24 && 0.40  & n.a && 0.40 & n.a \\
			{$AA^{T}$ \hfill }& 0.24 && 0.41  & n.a && 0.39 & n.a  \\
			\cmidrule{1-8} 
			{$TB$} & 0.20 && 0.41  & 0.02  && 0.40 & n.a \\
			{$DB$} & 0.12 && 0.39  & 0.018 && 0.40 & n.a  \\
			\bottomrule
		\end{tabular}
		\caption{Exploiting Matrix Properties: Lower Triangular ($L$), Symmetric Output, Tridiagonal ($T$) and Diagonal ($D$). Execution time (in sec) for $n=3000$.}
		\label{tab:exp3}
	\end{center}
\end{table}

\textbf{\textit{Test expressions:}} We consider the matrix multiplication
\begin{equation}
\label{eq:matmul}
Y := AB \qquad A,B \in \mathbb{R}^{n \times n}.
\end{equation}
The computational cost of the above expression using the GEMM kernel is approximately $2n^3$ FLOPs. The required number
of FLOPs can be lowered if the matrices in Expression~\ref{eq:matmul} have the following special properties (we report the execution times in Table~\ref{tab:exp3}):
\begin{itemize}
	\setlength\itemsep{1em}
	\item \textbf{\textit{Triangular:}} When all elements above the diagonal of $A$ are zero, then the matrix is
	lower-triangular; for clarity, we refer to it as $L$. 
	The matrix product $LB$ can be performed using the kernel TRMM for just $n^3$ FLOPs, which is half the cost of GEMM.
	\paragraph*{\textbf{Result}} 
	Presently, neither frameworks offer specialized routines that a user can explicitly invoke to take advantage of the triangular matrix structure in multiplication.         
	With the regular matrix multiplication operator, the execution time for $LB$  is close to $1.7\times$ higher than the corresponding SciPy implementation that explicitly calls TRMM.  Therefore, the frameworks do not offer provision to save the unnecessary computations.

	\item \textbf{\textbf{Symmetric output:}} In Expression~\ref{eq:matmul}, if $B$ happens to be $A^T$, then the output of $AA^T$ is a symmetric matrix. BLAS offers a
	specialized routine, SYRK (``SYmmetric Rank-K update''), which computes $AA^T$ with only $n^3$ FLOPs, which is half
	the FLOPs of a GEMM.
	\paragraph*{\textbf{Result}} The execution of $AA^T$ using SYRK through SciPy is again close to $1.7\times$ faster than both TF and PyT; the frameworks do not take advantage of this special structure to accelerate the computation.
\end{itemize} 

When the matrices in Expression~\ref{eq:matmul} have the following properties, the matrix product cannot be implemented
using a single specialized BLAS kernel, but can be effectively decomposed into a sequence of kernels: 
\begin{itemize}
	\setlength\itemsep{1em}
	
	\item \textbf{\textit{Tridiagonal:}} When all the elements except those in the three central diagonals of $A$ are zero, the matrix is Tridiagonal $T$.  The matrix product $TB$ can be reduced from $\mathcal{O}(n^3)$ to $\mathcal{O}(n^2)$ complexity by re-writing it as a sequence of scaling operations (using the SCAL kernel) applied to every row of $B$; the overall computation requires only $6n^2$ FLOPs.
	\paragraph*{\textbf{Result}}
	The execution of $TB$ using a sequence of SCAL via SciPy is $2\times$ faster than the regular matrix
	multiplication. While both the frameworks do not automatically exploit the tridiagonal structure of the
	matrix, TensorFlow does however offer a specialized method \textit{linalg.tridiagonal\_matmul} which the
	users themselves can manually invoke.  The execution through this optimized call is significantly faster than even the SciPy implementation, as TensorFlow takes advantage of the fact that the scaling operations can be executed simultaneously, and hence parallelizes them.

	\item \textbf{\textit{Diagonal:}}
	The diagonal matrix $D$ is a special case of the tridiagonal matrix $T$, in which, instead of the three central diagonals, only the main diagonal is non-zero.  The product $DB$ requires only $n^2$ FLOPs.
	
	\paragraph*{\textbf{Result}} The execution time when using the regular matrix multiplication is $4\times$ higher than the optimized SciPy implementation. TensorFlows optimized function \textit{linalg.tridiagonal\_matmul} is $10\times$ faster than SciPy.

\end{itemize}

\textbf{\textit{Discussion:}}  If the matrix properties are known before run-time,  users can explicitly annotate matrices with types that encode the properties, making it possible for the compiler to map the corresponding operations with those matrices to the specialized kernels; languages such as Julia~\cite{bezanson2017julia} already offer this feature. The compilers in TF and PyT could also exploit the optimized kernels if matrix properties are annotated on the frameworks computational graphs. The propagation of matrix properties through the graph would also facilitate algebraic simplifications that might speed up computations; for instance, if it is known that certain matrix $Q \in \mathbb{R}^{n \times n}$ is orthogonal, then it would be straightforward to deduce that the operation $Q^TQ$ would evaluate to an Identity matrix and avoid the explicit multiplication, hence saving $2n^3$ FLOPs.

\subsection{\textbf{Experiment 4: Algebraic Manipulation}}
Algebraic properties can be used to rewrite an expression in several alternative ways. Although these alternatives (or
variants) are mathematically equivalent, the number of FLOPs required can be vastly different.  In
Section~\ref{sec:introduction}, we illustrated the differences in FLOP counts as a result of rewriting
Expression~\ref{eqn:img-rest} using the distributive property. 
In this experiment, we develop benchmarks to expose some common algebraic manipulations that can result in significant performance improvements.

\begin{itemize}
	\setlength\itemsep{1em}
	\item \textbf{\textit{Distributivity:}} In general, the application of distributive property to an expression modifies the number of multiplication operations required; for instance, consider the following test expression, 
	\begin{equation}
	\label{eq:dist}
	AB+AC = A(B + C) \qquad A,B,C \in \mathbb{R}^{n \times n}
	\end{equation}
	The Left-Hand-side (LHS) of  Equation~\ref{eq:dist} requires two matrix multiplication operations, while the Right-Hand-Side (RHS) requires only one, which as a result saves $2n^{3}$ FLOPs. 
	However, the reduction in multiplication operations does not always reduce the required FLOPs; for instance, consider the following equation,
	\begin{equation}
	\label{eq:dist2}
	A\mathbf{x} - H^{T}(H\mathbf{x}) = (A - H^{T}H)\mathbf{x} \qquad A,H \in \mathbb{R}^{n \times n}, \mathbf{x} \in \mathbb{R}^{n}
	\end{equation}
	The  evaluation of LHS consist of three multiplication operations, i.e., three matrix-vector products of $\mathcal{O}(n^2)$ complexity, whereas the RHS, which although consist of only two multiplications, one of them is an expensive $\mathcal{O}(n^3)$ matrix-matrix multiplication; as a consequence, RHS requires more FLOPs than LHS despite the reduction in number of multiplications.
	\paragraph*{\textbf{Result}}The execution times for the LHS and RHS of the test equations are shown in Table~\ref{tab:exp4}. In both frameworks, the execution of the LHS of Equation~\ref{eq:dist} is approximately $2 \times$ longer than the RHS, and the RHS of Equation~\ref{eq:dist2} is around $40 \times$ longer than the LHS. As the execution times of LHS and RHS are not comparable, we infer that the frameworks do not rewrite expressions using distributivity to find a variant with fewer FLOPs. 
	
	
	\item \textbf{\textit{Blocked Matrices:}}  We consider the expression where matrices $A_1, A_2 \in \mathbb{R}^{\frac{n}{2} \times \frac{n}{2}}$  and $B_1, B_2 \in \mathbb{R}^{\frac{n}{2} \times n}$ are used to form the following large blocked matrix structures:
	\begin{equation*}
	A_B := \begin{bmatrix}
	\begin{matrix}
	A_1 & 0 \\  0 & A_2
	\end{matrix}
	\end{bmatrix} \in \mathbb{R}^{n \times n} \qquad
	B_B := \begin{bmatrix}
	\begin{matrix}
	B_1 \\  B_2
	\end{matrix}
	\end{bmatrix} \in \mathbb{R}^{n \times n}
	\end{equation*}
	Matrix products involving blocked matrices can be expressed as,
	\begin{equation}
	\label{eq:block}
	A_BB_B = 
	\begin{bmatrix}
	\begin{matrix}
	(A_1B_1) \\ (A_2B_2)
	\end{matrix}
	\end{bmatrix} 
	\end{equation}
	In many applications such as finite element methods~\cite{blatt2010parallel}, and signal processing~\cite{park2011concentration}, matrices exhibit such blocked structures.
	The matrix product $A_BB_B$ with the large matrices (LHS of Equation~\ref{eq:block}) is more expensive than the product with individual blocks (RHS of Equation~\ref{eq:block}). In Equation~\ref{eq:block}, while the evaluation of LHS costs $2n^3$ FLOPs, the RHS that consist of matrix multiplications with the small blocks, costs only $\frac{n^3}{2} + \frac{n^3}{2} =  n^3$ FLOPs.
	\paragraph*{\textbf{Result}} We construct the big matrix $A_B$ by explicitly concatenating the small matrices, so that the construction is captured in the frameworks computational graph. For $n=3000$, the  execution of the LHS of Equation~\ref{eq:block} is approximately $2 \times$ longer than the RHS (see Table~\ref{tab:exp4}). Therefore, we conclude that the frameworks do not exploit the blocked matrix structure to save computations.
	
\end{itemize}

\begin{table}[h!]
	\normalsize
	\begin{center}
		\renewcommand{\arraystretch}{1.2}
		\begin{tabular}{@{}l cc c cc@{}}
			\toprule
			Property & \multicolumn{2}{c}{TF} && \multicolumn{2}{c}{PyT} \\
			\cmidrule{2-3} \cmidrule{5-6}
			& LHS & RHS && LHS & RHS \\
			\cmidrule{1-6}
			{Distributivity Eq~[\ref{eq:dist}] \hfill }& 0.78  & 040 && 0.81 & 0.41  \\
			{Distributivity Eq~[\ref{eq:dist2}] \hfill }& 0.01  & 0.42 && 0.01 & 0.41  \\
			{Blocked matrices} & 0.40  & 0.20 && 0.40 & 0.20  \\
			\bottomrule
		\end{tabular}
		\caption{Algebraic Manipulations. Execution time (in sec) for $n=3000$.}
		\label{tab:exp4}
	\end{center}
\end{table}

\textbf{\textit{Discussion:}} Derivation graphs can be used to systematically rewrite and explore the variants of an input expression. 
The nodes of this graph represent intermediate expressions and the edges represent the operations and its associated cost to transform the expression. Starting from the original input expression, child nodes are added as long as there exist an operation that can simplify the expression from the parent node. Upon termination,  the different paths from root to leaves nodes are the alternative programs that compute the input problem and the total cost of a program can be obtained by summing the costs along the path. The program with minimum cost can be found by searching for the shortest path in the derivation graph. Linnea~\cite{barthels2021linnea} is a linear algebra code generator that uses such derivation graph to generate variants of input expressions and find optimal programs in terms of FLOPs. We remark that derivation graphs can serve as one of the top level intermediate representations in TF or PyT to facilitate algebraic optimizations.

\subsection{\textbf{\textit{Experiment 5: Code Motion}}}

In this final experiment, we test if the frameworks can identify operations that can be moved around or swapped with another to improve performance.

\begin{itemize}
	\setlength\itemsep{1em}
	\item \textbf{\textit{Loop-Invariant Code Motion:}} Operations that occur within a loop but yield the same result regardless of how many times the loop is executed, can be computed just once and moved outside the loop. The code snippet shown in Figure~\ref{fig:loop-inv} updates the matrix product $AB$ (where $A,B \in \mathbb{R}^{n \times n}$) with the outer products of the vectors $v_1, v_2, v_3 \in \mathbb{R}^n$.
	\begin{center}
		\includegraphics[width=0.45\textwidth]{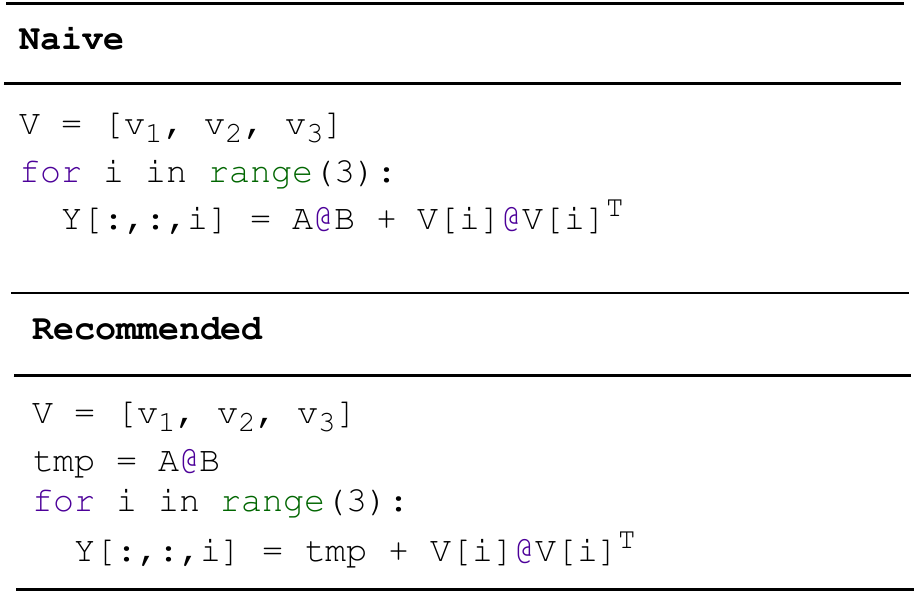}
		\captionof{figure}{Loop-Invariant code motion. }
		\label{fig:loop-inv}
	\end{center}
	In the ``naive'' implementation, the expensive $\mathcal{O}(n^3)$ operation $AB$ is recomputed in every iteration of the loop. However, the recommended implementation is to compute $AB$ only once, outside the loop body. 
	\paragraph*{\textbf{Result}} The execution times for the test expressions are shown in Table~\ref{tab:exp5}. 
	The frameworks exhibit comparable performance for both naive and recommended implementations. Therefore, we conclude
	that both the frameworks perform loop-invariant code motion to move redundant operations outside the body of the loop.
	
	\item \textbf{\textit{Partial Operand Access:}} In the code snippet shown in Figure~\ref{fig:partial-op}, the
	output requires only a single element of the matrix operations: $A+B$ and $AB$.       
	\begin{center}
		\includegraphics[width=0.45\textwidth]{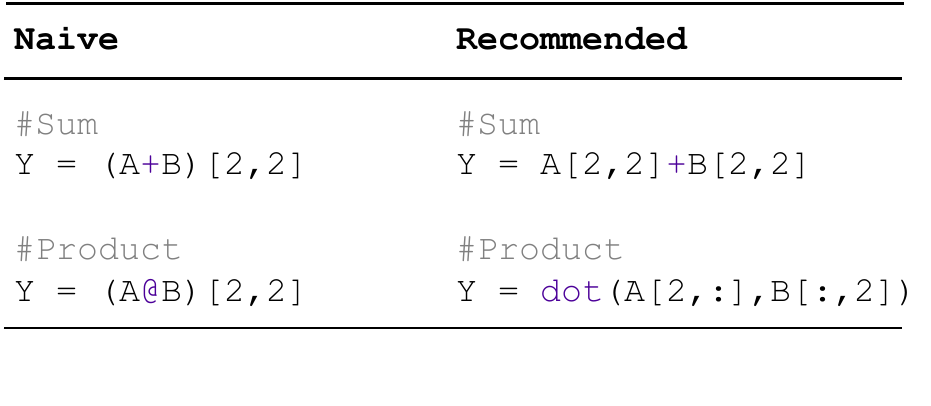}
		\captionof{figure}{Partial operand access.}
		\label{fig:partial-op}
	\end{center}
	The naive implementation consist of two operations: explicit matrix sum $A+B$ (or matrix product $AB$), followed by a slicing operation to access the required elements; however, the recommended implementation is to avoid the  $\mathcal{O}(n^2)$ matrix sum (or the  $\mathcal{O}(n^3)$ matrix product) by instead performing an $\mathcal{O}(1)$ element sum (or an $\mathcal{O}(n)$ dot product) of the required row and column elements (or slices) of $A$ and $B$ respectively. 
	\paragraph*{\textbf{Result}} In both frameworks, the execution time for the naive implementations are significantly higher than the recommended implementation\footnote{Recall from Sec.~\ref{sec:eval}, the decorator overheads for TF and PyT are 6e-4 sec and 2-e3 sec respectively.}. Hence, we infer that the frameworks do not swap the slicing and matrix operations to perform partial operand access.
\end{itemize}

\begin{table}[h!]
	\normalsize
	\begin{center}
		\renewcommand{\arraystretch}{1.2}
		\begin{tabular}{@{}l cc c cc@{}}
			\toprule
			Property & \multicolumn{2}{c}{TF} && \multicolumn{2}{c}{PyT} \\
			\cmidrule{2-3} \cmidrule{5-6}
			& Naive & Reco && Naive & Reco \\
			\cmidrule{1-6}
			{Loop-inv code motion \hfill }& 0.42  & 0.42 && 0.42 & 0.41 \\
			\cmidrule{1-6}
			{Partial-op access (sum)} & 0.011  & 6e-4 && 0.018 & 2e-3  \\
			{Partial-op access (product)} & 0.39  & 2e-3 && 0.40 & 3e-3  \\
			\bottomrule
		\end{tabular}
		\caption{Code Motion. Execution time (in sec) for $n=3000$.}
		\label{tab:exp5}
	\end{center}
\end{table}

\textbf{\textit{Discussion:}} 
The graph optimization systems of the frameworks are already known to perform graph re-writes that moves operations around to improve the performance for many neural network specific operations~\cite{larsen2019tensorflow,rotem2018glow}. 
Therefore, we remark that identifying code motions that speed up general linear algebra computations can further enhance the frameworks performance.

\section{Conclusion}
\label{sec:conclusion}

We considered the evaluation of linear algebra expressions, which are at the heart of countless machine learning problems. 
We focused on two of the most popular machine learning frameworks: TensorFlow and PyTorch, and developed experiments that expose opportunities for improving the performance when computing linear algebra expressions. 
Specifically, we analyzed and reported the performance gains that could be obtained through optimizations such as common sub-expression elimination, optimal parenthesization of matrix chains, applying the knowledge of matrix and linear algebra properties, and code motion. 
For each optimization, we presented guidelines to help both the framework developers and end-users achieve better performance.

The experiments included in our benchmarks are meant to expose an initial set of optimizations that we deemed essential to attaining high performance in linear algebra computations. 
Natural extension to this study include further investigation on other critical aspects of linear algebra computations such as exploitation of properties in the solution of linear systems, as well as the interplay of different optimizations, and taking into consideration the performance impacts due to the use of parallelism and accelerators. 

\bibliographystyle{IEEEtran}
\bibliography{lamp-ml}
\end{document}